\documentclass[11pt]{article}
\usepackage[T1]{fontenc}
\usepackage[final]{epsfig}
\usepackage{amssymb}
\usepackage{multicol}
\usepackage{graphics}
\usepackage{color}
\usepackage{ntimes}
\usepackage{graphicx}

\newcommand{\Tbar}{{\overline{T}}}
\newcommand{\nbar}{{\overline{n}}}

\newcommand{\BE}{\begin{equation}}
\newcommand{\EE}{\end{equation}}

\begin{document}

\title{Statistical variances in traffic data}

\author{Milan Krb\'alek${}^{1,2}$, Petr \v Seba${}^{2,3,4}$\\
\footnotesize{${}^1$ \emph{Faculty of Nuclear Sciences and
Physical Engineering, Czech Technical University, Prague - Czech
Republic}}\\  \footnotesize{${}^2$ \emph{Doppler Institute for
Mathematical Physics and Applied Mathematics,}}\\
\footnotesize{\emph{Faculty of Nuclear Sciences and Physical
Engineering,
Czech Technical University, Prague - Czech Republic}}\\
\footnotesize{${}^3$ \emph{University of Hradec Kr\'alov\'e,
Hradec
Kr\'alov\'e - Czech Republic}}\\
\footnotesize{${}^4$ \emph{Institute of Physics, Academy of
Sciences of the Czech Republic, Prague - Czech Republic}}}

\normalsize

\maketitle

\noindent \underline{\hspace{17cm}}

\begin{abstract} \small
We perform statistical analysis of the single-vehicle data
measured on the Dutch freeway A9 and discussed in
\cite{Dirk_and_Martin}. Using tools originating from the Random
Matrix Theory we show that the significant changes in the
statistics of the traffic data can be explained applying
equilibrium statistical physics of
interacting particles. \\


\end{abstract}

\noindent  \underline{\hspace{17cm}}\\

The detailed understanding of the processes acting in the traffic
systems is one of the most essential parts of the traffic
research. The basic knowledge of the vehicular interactions can be
found by means of the statistical analysis of the single-vehicle
data. As reported in Ref. \cite{Dirk}, \cite{Gas}, and \cite{HKT}
the microscopical traffic structure can be described with the help
of a repulsive potential describing the mutual interaction between
successive cars in the chain of vehicles. Especially, the
probability density $P_\beta(r)$ for the distance $r$ of the two
subsequent cars (\emph{clearance distribution}) can be described
with the help of an one-dimensional gas  having an inverse
temperature $\beta$ and interacting by a repulsive potential
$V(r)=r^{-1}$ (as discussed in Ref. \cite{Gas} and \cite{Dirk}).
Such a potential leads to a clearance distribution
\BE P_\beta(r)=A\, e^{-\frac{\beta}{r}-Br}, \label{clearance} \EE
where the constants $A=A(\beta),$ $B=B(\beta)$ fix up the proper
normalization $\int_0^\infty P_\beta(r)~dr=1$ and scaling
$\int_0^\infty rP_\beta(r)~dr=1.$ This distribution is in an
excellent agreement with the clearance distribution of real-road
data whereas the inverse temperature $\beta$  is related to the
traffic density $\varrho$.\\

Another way to seek for the interaction between  cars within the
highway data is to investigate the traffic flow fluctuations. One
possibility is to use  the so-called \emph{time-gap variance}
$\Delta_T$ considered in paper \cite{Dirk_and_Martin} and defined
as follows. Let $\{t_i:i=1 \ldots Q\}$ be the data set of time
intervals between subsequent cars passing a fixed point on the
highway. Using it one can calculate the moving average
$$T_k^{(N)}=\frac{1}{N} \sum_{i=k}^{k+N-1} t_i \hspace{0.5cm} (k=1\ldots Q-N+1)$$
of the time intervals produced by the $N+1$ successive vehicles
(i.e. $N$ gaps)
as well as the total average
$$\Tbar=\frac{1}{Q}\sum_{i=1}^Q t_i\equiv T_1^{(Q)}.$$
The time-gap variance $\Delta_T$ is defined by the variance of the
sample-averaged time intervals as a function of the sampling size
$N,$
$$\Delta_T=\frac{1}{Q-N+1}\sum_{k=1}^{Q-N+1} \left(T_k^{(N)}-\Tbar\right)^2,$$
where $k$ runs over all possible samples of $N+1$ successive cars.
For time intervals $t_i$ being statistically independent the law
of large numbers gives $\Delta_T(N)\propto 1/N$.\\

A statistical analysis of the data set recorded on the Dutch
freeway A9 and published in Ref. \cite{Dirk_and_Martin} leads,
however, to different results - see the Figure 1. For the free
traffic flow $(\varrho< 15~\mathrm{veh/km/lane})$ one observes
indeed the expected behavior $\Delta_T(N)\propto 1/N$. More
interesting behavior, nevertheless, is detected for higher
densities $(\varrho> 35~ \mathrm{veh/km/lane}).$ Here Nishinari,
Treiber, and Helbing (in Ref. \cite{Dirk_and_Martin}) have
empirically found a power law dependence
$$\Delta_T(N)\propto N^\gamma$$ with an exponent $\gamma \approx
-2/3,$ which can be explained as a manifestation of correlations
between the queued vehicles in a congested traffic flow.\\

\begin{figure} \centering
\scalebox{.5}{\includegraphics{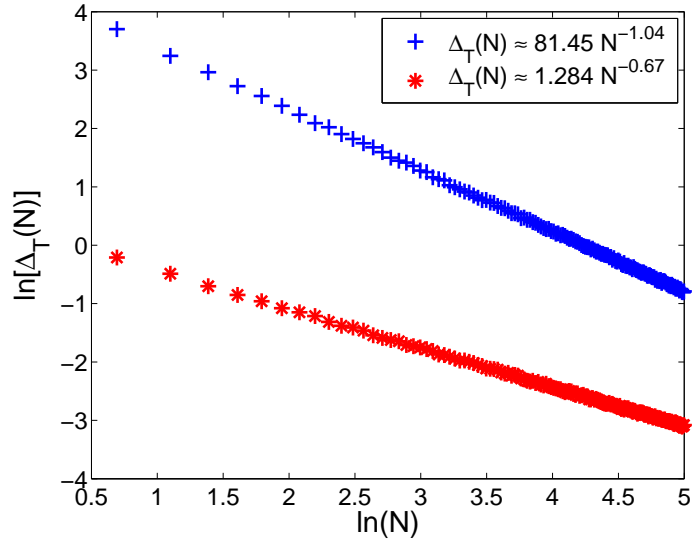}}
\parbox{12cm}{\caption{ \small The time-gap variance $\Delta_T(N)$ as
a function of the sampling size $N$ (in log-log scale). Plus signs
and stars represent the variance of average time-gaps for free and
congested flows, respectively.}} \label{fig:1}
\end{figure}

There is, however, one substantial drawback of this description.
The time-gap variance was introduced \emph{ad hoc} and hardly
anything is known about its exact mathematical properties in the
case of interacting vehicles. It is therefore appropriate to look
for an alternative that is mathematically well understood.  A
natural candidate is the \emph{number variance} $\Delta_n(L)$ that
was originally introduced for describing the statistics of
eigenvalues in the Random Matrix Theory. It reproduces also the
variances in the particle positions of certain class of
one-dimensional interacting gases (for example Dyson gas in Ref.
\cite{Mehta}).\\

Consider a set $\{r_i:i=1 \ldots Q\}$ of  distances (i.e.
\emph{clearances} in the traffic terminology) between each pair of
cars moving in the same lane. We suppose that the mean distance
taken over the complete set is re-scaled to one, i.e.
$$\sum_{i=1}^Q r_i=Q.$$
Dividing the interval $[0,Q]$ into subintervals $[(k-1)L,kL]$  of
a length $L$ and denoting by $n_k(L)$ the number of cars in the
$k$th subinterval, the average value $\nbar(L)$ taken over all
possible subintervals is
$$\nbar(L)=\frac{1}{\lfloor Q/L \rfloor} \sum_{k=1}^{\lfloor Q/L
\rfloor} n_k(L)=L,$$
where the integer part $\lfloor Q/L \rfloor$ stands for the number
of all subintervals $[(k-1)L,kL]$ included in the interval
$[0,Q].$ Number variance $\Delta_n(L)$ is then defined as
$$\Delta_n(L)=\frac{1}{\lfloor Q/L \rfloor} \sum_{k=1}^{\lfloor Q/L
\rfloor} \left(n_k(L)-L\right)^2$$
and represents the statistical variance of the number of vehicles
moving at the same time inside a fixed part of the road of a
length $L.$ The mathematical properties of the number variance are
well understood. For independent events one gets $\Delta_n(L)=L$.
Applying it to the highway data in the low density regime (free
traffic) one obtains however $\Delta_n(L)\approx 5L/6$ (not
plotted). The small deviation from the behavior $\Delta_n(L)=L$ is
induced by the weak (but still nonzero) interaction among the
cars.\\

The situation becomes more tricky when a congested traffic is
investigated. The touchy point is that behavior of the number
variance is sensitive to the temperature of the underlying gas -
or in the terminology of the Random Matrix Theory - to the
universality class of the random matrix ensemble. To use the known
mathematical results one has not to mix together states with
different densities - a procedure known as \emph{data unfolding}
in the Random Matrix Theory. For the transportation this means
than one cannot mix together traffic states with different traffic
densities and hence with different vigilance of the drivers. So we
will perform a separate analysis of the data-samples lying within
short density intervals to prevent so the undesirable mixing of
the different states.\\

We divide the region of the measured densities $\varrho \in
[0,85~\mathrm{veh/km/lane}]$ into eighty five equidistant
subintervals and  analyze the data from each one of them
separately. The number variance $\Delta_n(L)$ evaluated with the
data in a fixed density interval has a characteristic linear tail
(see Fig. 2) that is well known from the Random Matrix Theory.
Similarly, such a behavior was found in models of one-dimensional
thermodynamical gases with the nearest-neighbor repulsion among
the particles (see Ref. \cite{Bogomolny}). We remind that for the
case where the interaction is not restricted to the nearest
neighbors but includes all particles the number variance has
typically a logarithmical tail - see \cite{Mehta}. So the linear
tail of $\Delta_n(L)$  supports the view that in the traffic
stream the interactions are restricted to the few nearest cars
only. The slope of the linear tail of  $\Delta_n(L)$ decreases
with the traffic density (see the top subplot in the Fig. 3). It
is a consequence of the increasing alertness of the drivers and
hence of the increasing coupling between the neighboring cars in
the
dense traffic flows.\\

The fact that the behavior of the number variance evaluated from
the traffic data coincides with the results obtained for
interacting one-dimensional gases strengthen the idea to apply the
equilibrium statistical physics for describing the local
properties of the traffic flow. We take the advantage of this
approach in a following thermodynamical traffic model.\\

Consider $N$ identical particles (cars) on a circle of the
circumference $N$ exposed to the thermal bath with inverse
temperature $\beta.$ Let $x_i$ $(i=1 \ldots N)$ denote the
position of the $i$-th particle and put $x_{N+1}=x_1+N,$ for
convenience. The particle interaction is described  by a potential
(see Ref. \cite{Dirk})
\BE U \propto \sum_{i=1}^N r_i^{-1}, \label{power-law-potential}
\EE
where $r_i=|x_{i+1}-x_i|$ is the distance between the neighboring
particles. The nearest-neighbor interaction is chosen with the
respect to the realistic behavior of a car-driver in the traffic
stream. As published in Ref. \cite{Gas}, the heat bath drives the
model into the thermal equilibrium and the probability density
$P_\beta(r)$ for gap $r$ among the neighboring particles
corresponds to the function (\ref{clearance}).\\

\begin{figure} \centering
\scalebox{.5}{\includegraphics{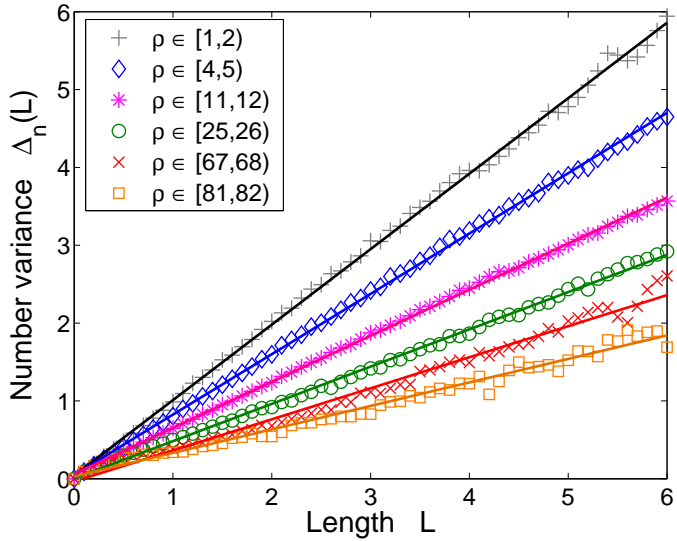}}
\parbox{12cm}{\caption{ \small Number variance $\Delta_n(L)$ as a function of
length $L.$ Plus signs, diamonds, stars, circles, crosses, and
squares represent the number variance of real-road data in
selected density regions (see legend for details). The curves show
the linear approximations of the individual data. Their slopes
were carefully analyzed and consecutively visualized on the Fig. 3
(top part).}} \label{fig:2}
\end{figure}

According to \cite{Mehta}, the number variance $\Delta_n(L)$ of an
one-dimensional gas in thermal equilibrium can be exactly
determined from its spacing distribution $P_\beta(r).$  For the
clearance distribution (\ref{clearance}) we obtain (see
\cite{unpublished})
\BE \Delta_n(L)\approx\chi L+\gamma \hspace{1cm} (L\rightarrow
\infty),\label{NV_exact}\EE
i.e. a linear function with a slope
$$\chi\approx 1+\frac{3-16B\beta-16\sqrt{B\beta}}{\left(3+4\sqrt{B\beta}\right)^2}$$
and
$$\gamma\approx\frac{32B\beta\left(21+8B\beta+24\sqrt{B\beta}\right)-48\sqrt{B\beta}-63}{24B\left(3+4\sqrt{B\beta}\right)^2},$$
which depend on the inverse temperature $\beta$ only. Above
relations represent a large $\beta$ approximations whereas two
phenomenological formulae
$$\chi \approx \frac{1}{2.4360~\beta^{0.8207}+1}$$
and
$$\gamma \approx \frac{\beta}{5.1926~\beta+2.3929}$$
specify the behavior of $\chi$ and $\gamma$  near the origin. We
emphasize that, in the limiting  case $\beta=0,$ the value of
$\chi$ is equal to one, i.e.  $\Delta_n(L)=L,$ as expected for the
independent events. The slope $\chi$ is a  decreasing function of
$\beta.$\\

\begin{figure} \centering
\scalebox{.5}{\includegraphics{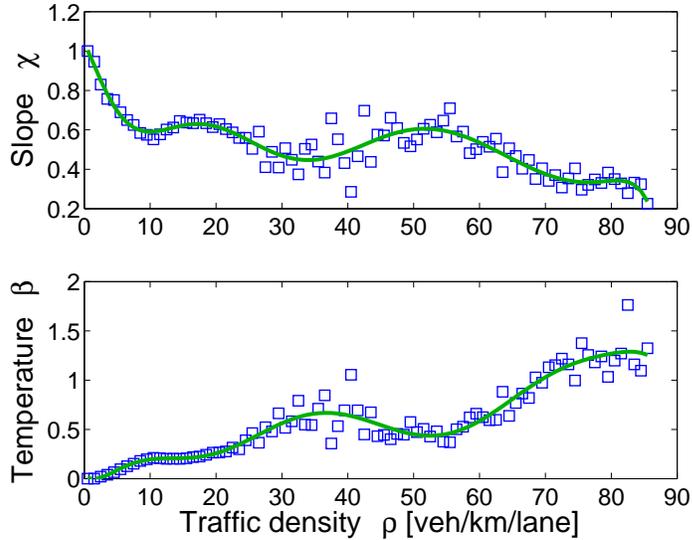}}
\parbox{12cm}{\caption{ \small The slope $\chi$ and the inverse temperature
$\beta$ as functions of the traffic density $\varrho.$ The squares
on the upper subplot display the slope of the number variance
$\Delta_n(L)$ (see Fig. 2), separately analyzed for various
traffic densities. The lower subplot visualizes the fitted values
of the inverse temperature $\beta,$ for which the exact form of
number variance $\Delta_n(L)=\chi(\beta)~L+\gamma(\beta)$
corresponds to the number variance obtained from the traffic data.
The continuous curves represent a polynomial approximations of the
relevant data.}} \label{fig:3}
\end{figure}

The described properties of the function $\Delta_n(L)$ agree with
the behavior of the number variance extracted from the traffic
data (see Fig. 2). A comparison between traffic data number
variance and the formula (\ref{NV_exact}) allows us to determine
the empirical dependence of inverse temperature $\beta$ on traffic
density $\varrho$. The inverse temperature reflects the
microscopic status in which the individual vehicular interactions
influence the traffic.  Conversely, in the macroscopic approach,
traffic is treated as a continuum and modelled by aggregated,
fluid-like quantities, such as density and flow (see
\cite{Helbing}). Its most prominent result is the dependence of
the
traffic flow on the traffic density - the fundamental diagram.\\

It is clear that the macroscopic traffic characteristics are
determined by its microscopic status. Consequently there should be
a relation between the behavior of the fundamental diagram and
that of the inverse temperature $\beta$. On the Figure 4 we
display the behavior of  the inverse temperature $\beta$
simultaneously with the fundamental diagram. The both curves show
a virtually linear increase in the region of  a free traffic (up
to $\varrho \approx 10~\mathrm{veh/km/lane}$). The inverse
temperature $\beta$ then displays a plateau for the densities up
to $18~\mathrm{veh/km/lane}$ while the flow continues to increase.
A detailed inspection uncovers, however, that the increase of the
traffic flow ceases to be linear and becomes concave at that
region. So the flow is reduced with respect to the outcome
expected for a linear behavior - a manifestation of the onset of
the phenomenon that finally leads to a congested traffic. For
larger densities the temperature $\beta$ increases up to $\varrho
\gtrapprox 32~\mathrm{veh/km/lane}$. The center of this interval
is localized at $\varrho \approx 25 $ - a critical point of the
fundamental diagram at which the flow starts to decrease. This
behavior of the inverse temperature is understandable and imposed
by the fact that the drivers, moving quite fast in a relatively
dense traffic flow, have to synchronize their driving with the
preceding car (a strong interaction) and are therefore under a
considerable psychological pressure. After the transition from the
free to a congested traffic regime (between 40 and
$50~\mathrm{veh/km/lane}$), the synchronization continues to
decline because of the decrease in the mean velocity leading to
decreasing $\beta$. Finally - for densities related to the
congested traffic - the inverse temperature increases while the
flow remains constant. The comparison between the traffic flow and
the inverse temperature is even more illustrative when the changes
of the flow are taken into account. Therefore we evaluate the
derivative of the flow
$$J'=\frac{\partial J}{\partial \varrho}$$
and plot the result on the Figure 5. The behavior of the inverse
temperature $\beta$ can be understood as a quantitative
description of the alertness required by the
drivers in a given situation.\\

\begin{figure} \centering
\scalebox{.5}{\includegraphics{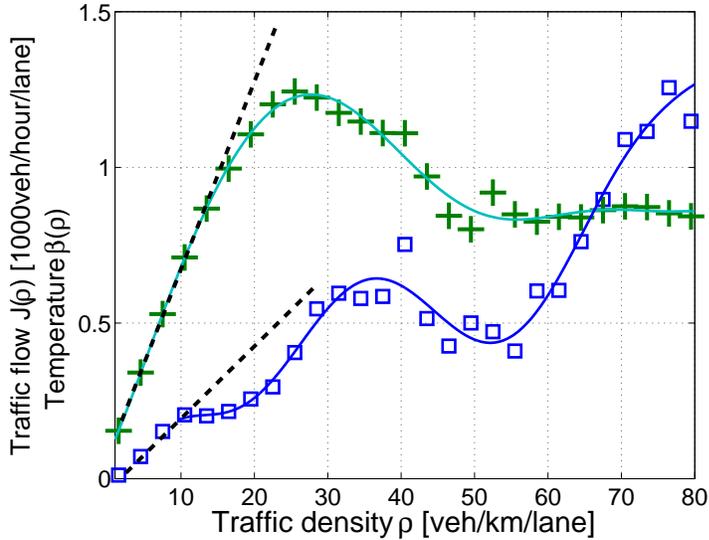}}
\parbox{12cm}{\caption{ \small Traffic flow $J(\varrho)$ and inverse temperature
$\beta(\varrho)$ as functions of a traffic density $\varrho$. Plus
signs display a traffic flow in thousands of vehicles per hour and
squares correspond to inverse temperature of the traffic gas. The
results of a polynomial curve-fitting are visualized by the
continuous curves. The dashed lines represent a linear
approximations of the relevant data near the origin.}}
\label{fig:4}
\end{figure}

The dependence of $\beta$ on the density $\varrho$ can be obtained
also using the measured clearance distribution and comparing it
with the formula (\ref{clearance}). It leads to the same results
as $\beta$ obtained from the number variance $\Delta_n(L)$. It is
known (see \cite{Mehta}) that for one-dimensional gases in thermal
equilibrium the function $\Delta_n(L)$  can be determined from the
knowledge of the spacing distribution $P_\beta(r).$ So the fact
that obtaining $\beta$ by virtue of the number variance
$\Delta_n(L)$ and the spacing distribution $P_\beta(r)$ leads to
the same results supports the view that locally the traffic can be
described by
instruments of equilibrium statistical physics.\\

\begin{figure} \centering
\scalebox{.5}{\includegraphics{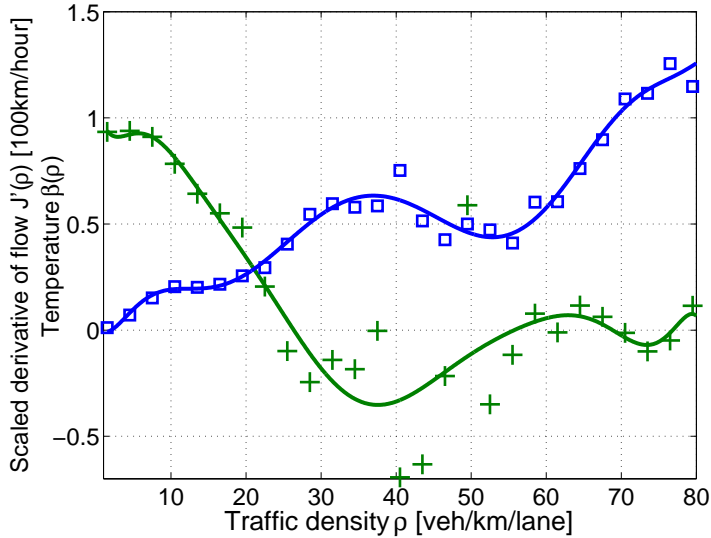}}
\parbox{12cm}{\caption{ \small Inverse temperature $\beta(\varrho)$ and scaled derivative
of the flow $J'=\frac{\partial J}{\partial \varrho}$ as functions
of a traffic density $\varrho.$ Squares correspond to the inverse
temperature of traffic sample while plus signs display the first
derivative of the flow (scaled for better visualization). The
continuous curves represent a relevant polynomial
approximations.}} \label{fig:5}
\end{figure}

In summary, we have investigated the statistical variances of the
single-vehicle data from the Dutch freeway A9. Particularly,  we
have separately analyzed the number variance in eighty five
equidistant density-subregions and found a linear dependence in
each of them. Using the thermodynamical model presented originally
in Ref.\cite{Dirk}, we have found an excellent agreement between
the number variance of particles in thermal-equilibrium and that
of the traffic  data. It was demonstrated that the inverse
temperature of the traffic sample, indicating the degree of
alertness of the drivers, shows an increase at both the low and
high densities. In the intermediate region, where the free flow
regime converts to the
congested traffic, it displays more complex behavior.\\

The presented results support the possibility for applying the
equilibrium statistical physics to the traffic systems. It
confirms also the hypothesis that short-ranged forwardly-directed
power-law potential (\ref{power-law-potential}) is a good choice
for describing the fundamental interaction among
the vehicles.\\

\noindent \Large{\textbf{Acknowledgements}}\\
\normalsize

We would like to thank Dutch Ministry of Transport for providing
the single-vehicle induction-loop-detector data. This work was
supported by the Ministry of Education, Youth and Sports of the
Czech Republic within the projects LC06002 and MSM 6840770039.

\end{document}